\documentclass[10pt, twocolumn, aps, nofootinbib, longbibliography, nobibnotes]{revtex4-1}
\pdfoutput=1
\usepackage{amsfonts, amssymb, amsmath, amsthm}
\usepackage{multirow}
\usepackage{latexsym}
\usepackage{pbox}
\usepackage[tracking=smallcaps]{microtype}	
\usepackage{url}
\usepackage{color}
\definecolor{DarkGray}{rgb}{0.1,0.1,0.5}
\usepackage[colorlinks=true,breaklinks, linkcolor=DarkGray,citecolor=DarkGray,urlcolor=DarkGray]{hyperref}	

\usepackage{graphicx}
\usepackage[caption=false]{subfig}

\newcommand{\bra}[1]{{\langle#1|}}
\newcommand{\ket}[1]{{|#1\rangle}}
\newcommand{\braket}[2]{{\langle#1|#2\rangle}}
\newcommand{\ketbra}[2]{{\ket{#1}\!\bra{#2}}}

\newcommand{\abs}[1]{{\lvert #1\rvert}}	

\def\A {{\mathcal A}}

\def\H {{\mathcal H}}

\def\cP {{\mathcal P}}

\def\W {{\mathcal W}}

\newcommand{\identity}{\ensuremath{\boldsymbol{1}}} 
\newcommand{\Id}{\identity}

\newcounter{sprows}

\newlength{\spheight}
\newlength{\spraise}

\newlength{\commentslength}

\newcommand{\rem}[1]{}

\newfont{\subsubsecfnt}{ptmri8t at 11pt}
\renewcommand{\subparagraph}[1]{\smallskip{\subsubsecfnt #1.}}

\newcommand{\eqnref}[1]{\hyperref[#1]{{(\ref*{#1})}}}
\newcommand{\thmref}[1]{\hyperref[#1]{{Theorem~\ref*{#1}}}}
\newcommand{\lemref}[1]{\hyperref[#1]{{Lemma~\ref*{#1}}}}
\newcommand{\corref}[1]{\hyperref[#1]{{Corollary~\ref*{#1}}}}
\newcommand{\defref}[1]{\hyperref[#1]{{Definition~\ref*{#1}}}}
\newcommand{\secref}[1]{\hyperref[#1]{{Sec.~\ref*{#1}}}}
\newcommand{\figref}[1]{\hyperref[#1]{{Fig.~\ref*{#1}}}}  
\newcommand{\tabref}[1]{\hyperref[#1]{{Table~\ref*{#1}}}}
\newcommand{\remref}[1]{\hyperref[#1]{{Remark~\ref*{#1}}}}
\newcommand{\appref}[1]{\hyperref[#1]{{Appendix~\ref*{#1}}}}
\newcommand{\claimref}[1]{\hyperref[#1]{{Claim~\ref*{#1}}}}
\newcommand{\factref}[1]{\hyperref[#1]{{Fact~\ref*{#1}}}}
\newcommand{\propref}[1]{\hyperref[#1]{{Proposition~\ref*{#1}}}}
\newcommand{\exampleref}[1]{\hyperref[#1]{{Example~\ref*{#1}}}}
\newcommand{\conjref}[1]{\hyperref[#1]{{Conjecture~\ref*{#1}}}}

\allowdisplaybreaks[1]

\def\COLOR{}
\ifdefined\COLOR

\else

\fi

\definecolor{Cayenne}{rgb}{0.5,0,0}
\definecolor{Midnight}{rgb}{0,0,0.5}
\definecolor{Plum}{rgb}{0.5,0,0.5}
\definecolor{Teal}{rgb}{0,0.5,0.5}
\definecolor{Clover}{rgb}{0,0.5,0}
\definecolor{Maroon}{rgb}{0.5,0,0.25}
\definecolor{Ocean}{rgb}{0,0.25,0.5}
\definecolor{Tangerine}{rgb}{1,0.5,0}
\definecolor{Strawberry}{rgb}{1,0,0.5}
\definecolor{Fern}{rgb}{0.25,0.5,0}
\definecolor{Aqua}{rgb}{0,0.5,1}
\definecolor{Moss}{rgb}{0,0.5,0.25}
\definecolor{Mocha}{rgb}{0.5,0.25,0}
\definecolor{Lemon}{rgb}{1,1,0}
\definecolor{Asparagus}{rgb}{0.5,0.5,0}
\definecolor{Grape}{rgb}{0.5,0,1}
\definecolor{Iron}{rgb}{.3,.3,.3}
\definecolor{Steel}{rgb}{.4,.4,.4}

\usepackage{soul} 
\usepackage{enumitem} 
\usepackage{listings}
\usepackage{tabularx}
\usepackage{lipsum}
\usepackage{hyperref}

\begin{document}
\def\compilefullpaper{}

\title{Influence of coin symmetry on infinite hitting times in quantum walks}
\author{Prithviraj Prabhu and Todd A. Brun}
\affiliation{University of Southern California}

\begin{abstract}

Classical random walks on finite graphs have an underrated property: a walk from any vertex can reach every other vertex in finite time, provided they are connected.  Discrete-time quantum walks on finite connected graphs however, can have infinite hitting times.  This phenomenon is related to graph symmetry, as previously characterized by the group of direction-preserving graph automorphisms that trivially affect the coin Hilbert space.  If a graph is symmetric enough (in a particular sense) then the associated quantum walk unitary will contain eigenvectors that do not overlap a set of target vertices, for any coin flip operator. These eigenvectors span the Infinite Hitting Time (IHT) subspace.  Quantum states in the IHT subspace never reach the target vertices, leading to infinite hitting times.  However, this is not the whole story: the graph of the $3$D cube does not satisfy this symmetry constraint, yet quantum walks on this graph with certain symmetric coins can exhibit infinite hitting times.  We study the effect of coin symmetry by analyzing the group of coin-permutation symmetries (CPS): graph automorphisms that act nontrivially on the coin Hilbert space but leave the coin operator invariant.  Unitaries using highly symmetric coins with large CPS groups, such as the permutation-invariant Grover coin, are associated with higher probabilities of never arriving, as a result of their larger IHT subspaces.  
\end{abstract}

\maketitle

\section{Introduction}

A classical random walk can be described by the movement of a particle on a graph, where the vertices represent possible locations and the edges connect neighboring sites.  At each step, the next edge is chosen randomly \cite{Pear05}.  Quantum walks are the unitary analogues of these classical random walks \cite{Venegas-Andraca2012, Kempe2009}.  Similar to the classical case, there are discrete-time \cite{Ambainis, Aharonov2001} and continuous-time \cite{Farhi1998, Childs2001} versions of quantum walks.  In this paper, we only consider discrete-time quantum walks.

The unitary operator $\hat U$ corresponding to one time step in the discrete-time quantum walk can be decomposed into two parts: $\hat U = \hat S(\hat \Id_v \otimes \hat C)$.  The Hilbert space of $\hat U$ is $\H = \H_v \otimes \H_c$, where $\H_v$ and $\H_c$ are the Hilbert spaces of the position (vertices) and of the internal degree of freedom (coin), respectively.  The shift operator $\hat S$ propagates quantum states between the vertices of the graph.  It is analogous to the adjacency matrix of the graph, but since the shift depends on the state of the coin it operates on the complete Hilbert space $\H$.  It takes the form
\begin{equation}
\hat{S} = \sum_{v,c} \ket{c(v)}\bra{v} \otimes \ket{c}\bra{c} ,
\end{equation}
where $\{\ket{v}\}$ are vertex basis states, $\{\ket{c}\}$ are coin basis states, and $\ket{c(v)}$ is the neighboring vertex to $v$ along the edge labeled by coin state $c$.

The coin operator $\hat C$ acts on the coin Hilbert space $\H_c$.  Discrete-time quantum walks use these operators to mix the particle's internal basis states at each vertex.  A vertex with $d$ edges (degree $d$) allows the particle to have as many coin basis states.  In irregular graphs, the degree is different for different vertices, which causes the dimension of the coin Hilbert space to occupy a range of values, and hence it becomes difficult to choose one coin to act on all the vertices.  However there are a couple of solutions one can employ.  Self-loops can be used to effect regularity.  Another solution is for the coin operator to depend on the vertex.  Hence $\hat C$ can be structured differently for different applications.  In this paper we will only consider regular graphs.

The notion of vertex hitting time is a concept borrowed from classical random walks.  Classically it is defined as the average number of time steps taken for a walk to reach a ``final'' vertex, $v_f$, given an initial distribution.  The formula for the hitting time $\tau_h(v_f)$ is
\begin{equation}
\tau_h(v_f) = \sum_{t = 0}^{\infty} t p(v_f,t),
\end{equation} 
where $p(v_f,t)$ is the probability that the walk has reached vertex $v_f$ at time $t$ for the first time.

Classical random walks will always have finite hitting times for all vertices in finite connected graphs.  However, interference effects may prevent this in quantum walks.  This leads to the uniquely quantum mechanical phenomenon of an infinite hitting time. 

To define a quantum notion of hitting time we must define what it means to arrive at a vertex for the first time.  In our approach, we treat the final vertex as an absorbing wall.  After every time step of the walk, a measurement is performed on the final vertex to check if the walk has reached it or not.  If it has, the walk is halted.  The average time until the walk halts is the hitting time for the measured quantum walk.  The projective measurements used are $\hat \Pi_f = \ket{v_f} \bra{v_f} \otimes \Id_c$ and $\Id - \hat \Pi_f$, where $\ket{v_f}$ is the state on the final vertex.  For a detailed explanation, we refer the reader to~\cite{Krovi2007}.

\begin{figure*}
\centering
\includegraphics[scale=0.6]{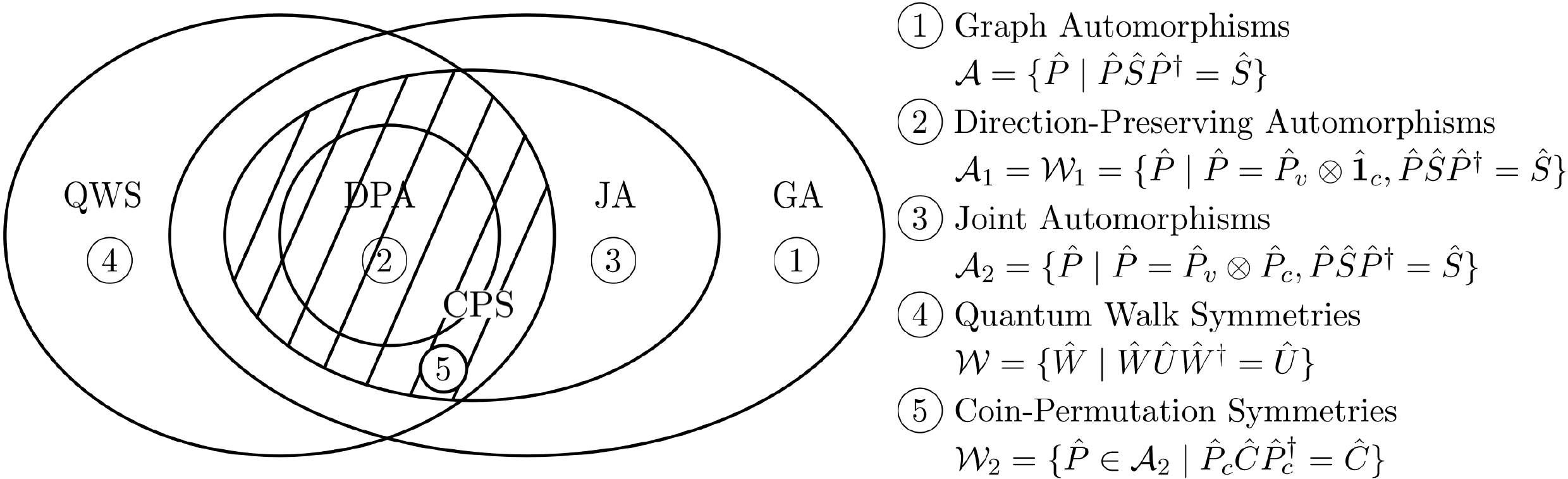}
\caption{Relations between classes of graph automorphisms and quantum walk symmetries.}
\label{f:symmetries}
\end{figure*}

Spectral analysis of the unitary evolution operator reveals that there may be some eigenvectors that have no overlap with the chosen final vertex $v_f$.  If a component of the initial state $\ket \psi$ exists in the subspace spanned by these eigenvectors, that component never reaches the final vertex. This phenomenon may lead to infinite hitting time at the final vertex, so we label this the Infinite Hitting Time (IHT) subspace, denoted as $V$.  The overlap between the initial state and the IHT subspace quantifies the probability that the final vertex is never reached:
\begin{equation}
O_{V}(\ket \psi) = \sum_{i = 1}^{|V|} \mid \braket{\psi}{V_i} \mid^2,
\label{e:ssIHT1corrected}
\end{equation}
where $\{V_i\}$ is an orthonormal basis for $V$. 

A sufficient condition for infinite hitting times, established in~\cite{Krovi2006a, Krovi2006, Krovi2007}, is for the evolution unitary to contain at least one $k$-dimensional eigenspace, where $k > d$.  Within this eigenspace there will exist a subspace of dimension at least $k-d$ that has no support on the final vertex.  This subspace forms a part of the IHT subspace. If multiple eigenspaces contain IHT subspaces, the sum of these subspaces gives the IHT subspace of the entire unitary. 

While the above condition is sufficient for a quantum walk unitary to have an IHT subspace, it is not a necessary condition.  It is possible for eigenspaces with degeneracy $\leq d$ to contain vectors that have no overlap with the final vertex.  Hence one must perform an analysis over all the eigenspaces when computing the size of the IHT subspace.  The procedure is simple.  Apply Gaussian elimination to each eigenspace and subsequently count the number of vectors that have no overlap with a chosen final vertex.

In general, degenerate eigenspaces of the unitary contribute to the existence of IHT subspaces. This degeneracy is primarily a consequence of symmetry, hence an analysis of the symmetries of the quantum walk may provide conditions predicting infinite hitting times.

\section{Symmetries of a quantum walk}

Previous work has focused on the influence of graph symmetry on infinite hitting time~\cite{Krovi2006, Krovi2006a}.  Here, the symmetry of the graph was represented by its group of direction-preserving automorphisms.  The irreducible representations of this group give a sufficient condition for a quantum walk on this graph to have infinite hitting times.  If the dimension of at least one of the irreducible representations is larger than the dimension of the coin, a corresponding eigenspace of the quantum walk unitary will have larger dimension than the coin, which is a sufficient condition for a graph to exhibit infinite hitting times. These papers considered Cayley graphs, where the automorphism group is simply related to the group used to define the graph.  The first discovery of infinite hitting time \cite{Krovi2006a} was observed on the $n$-dimensional hypercube with $2^n$ vertices, where $n\geq 3$.

An automorphism of a graph is a reordering of the vertices that leaves the graph unchanged. Since this reordering is just a permutation of the vertices, we can use permutation matrices to define automorphisms.  We consider two classes of permutations $\cP_1 \subseteq \cP_2 \subseteq \cP$ where $\cP$ is the set of all permutations on $\H$.  $\cP_1$ and $\cP_2$ contain permutations with a tensor product structure and are defined as:
\begin{equation}
\cP_1 = \{\hat P \mid \hat P = \hat P_v \otimes \hat \Id_c \},
\end{equation}
and
\begin{equation}
\cP_2 = \{\hat P \mid \hat P = \hat P_v \otimes \hat P_c \},
\end{equation}
where $\hat P_v$ and $\hat P_c$ are permutations acting on $\H_v$ and $\H_c$ respectively.  Permutations $\hat P$ that give rise to graph automorphisms necessarily leave the shift operator invariant:
\begin{equation}
\A = \{ \hat P \mid \hat P \hat S \hat P^\dagger = \hat S\} . 
\label{e:auto}
\end{equation}  

Thus the group of \textit{direction-preserving automorphisms} studied in~\cite{Krovi2006a, Krovi2006, Krovi2007}, is generated from permutations in $\cP_1$ that leave $\hat S$ invariant:
\begin{equation}
\A_1 = \{ \hat P \in \cP_1 \mid\hat P \hat S \hat P^\dagger = \hat S \}.
\end{equation}

Here, we consider automorphisms from the larger class of permutations $\cP_2$:
\begin{equation}
\A_2 = \{ \hat P \in \cP_2 \mid\hat P \hat S \hat P^\dagger = \hat S \},
\end{equation}
which forms the group of \textit{joint automorphisms}.  Clearly $\A_1 \subseteq \A_2 \subseteq \A$.  

Infinite hitting times in quantum walks are a consequence of symmetry in the quantum walk unitary ($\hat U$).  These symmetries can be represented by unitary operations $\hat W$ forming the group of quantum walk symmetries
\begin{equation}
\label{eq:walksymm}
\W = \{ \hat W \mid \hat W \hat U \hat W^\dagger = \hat U \} .
\end{equation}
Hence for a graph automorphism to be a symmetry of the quantum walk, it must satisfy $\hat P \hat U \hat P^\dagger = \hat U$.  This is readily satisfied by all the elements of $\A_1$.  Hence the group of direction-preserving automorphisms actually contains symmetries of the quantum walk: $\A_1 = \W_1$.  For elements in $\A_2$, $\hat P \hat U \hat P^\dagger = \hat U$ is only satisfied when the coin permutation leaves the coin operator invariant:
\begin{equation}
\label{eq:pc}
\hat P_c \hat C \hat P_c^\dagger = \hat C.
\end{equation}
The joint automorphisms that satisfy this condition are symmetries of the quantum walk and are denoted as the group of \textit{coin-permutation symmetries}:
\begin{equation}
\W_2 =\{ \hat P \in \A_2 \mid \hat P_c \hat C \hat P_c^\dagger = \hat C\}.
\end{equation}
Evidently, coin symmetry plays a large role in determining the size of the group of coin-permutation symmetries.  By choosing coins of varying symmetry, it may be possible to connect characteristics of the coin-permutation symmetry group to the existence of infinite hitting times.

Automorphisms of the graph from $\W \cap (\A \setminus \A_2)$ may still be symmetries of the quantum walk. We do not address symmetries of this type and leave it open to future research.  \figref{f:symmetries}  summarizes the relations between the different symmetry groups discussed above.

\section{Coins}

In the analysis of infinite hitting times in quantum walks, we use three coins.  On one end, we have the Grover coin which is the most symmetric of the lot. On the other end, we consider a random unitary, which should lack any particular symmetry.  As an intermediate between these two, we also consider the Discrete Fourier Transform (DFT) coin. 
\begin{itemize}
\item \textbf{Grover Coin}: This is the most symmetric coin generally used in quantum walks, and we would expect to see this symmetry reflected in the eigenspace decomposition of a walk's unitary.  The symmetry arises from the permutation invariance~\cite{Moore2002} of the coin which takes the form

\begin{equation}
\hat{G} =  \left[ {\begin{array}{ccccc}
	a 		& b 		& b 		& \cdots & b \\
	b 		& a 		& b 		& \cdots & b \\
	b 		& b		 	& a			& \cdots & b \\
	\vdots 	& \vdots 	& \vdots 	& \ddots & \vdots \\
	b 		& b			& b			& \cdots & a \\
															
\end{array} } \right],
\end{equation}  
where we choose $a = \frac{2}{d} -1$ and $b = \frac{2}{d}$ and $d$ is the dimension of the coin.  Due to the permutation invariance property, all $d!$ permutation matrices $\hat P_c$ of size $d \times d$ satisfy the coin-permutation symmetry constraint: $\hat P_c \hat G \hat P_c = \hat G $, implying this coin is associated with the largest possible coin-permutation symmetry group.

The symmetries of this coin fit into the symmetries of the hypercube very nicely.  A quantum walk on an $n$-dimensional hypercube with this coin can be reduced to a quantum walk on a line, for the initial state $\ketbra 0 0 \otimes \ketbra \psi \psi$ where $\ket \psi = \tfrac{1}{\sqrt{n}} \sum_{j=1}^{n} \ket j$~\cite{Shenvi2003}.

\item \textbf{DFT Coin}: Another commonly used coin is the DFT coin, $\hat D$. It is a generalization of the popular Hadamard coin, $\hat J$, to arbitrary dimension:

\begin{equation}
\hat{J} = \frac{1}{\sqrt{2}} \left[ {\begin{array}{cc}
										1 & 1 \\
										1 & -1\\
															
\end{array} } \right],
\end{equation}  

\begin{equation}
\hat{D} = \frac{1}{\sqrt{d}} \left[ {\begin{array}{ccccc}
	1 		& 1 			& 1 					& \cdots & 1 \\
	1 		& \omega 		& \omega^2 				& \cdots & \omega^{(d-1)} \\
	1 		& \omega^2 		& \omega^4 				& \cdots & \omega^{2 \cdot (d-1)} \\
	\vdots 	& \vdots 		& \vdots 				& \ddots & \vdots \\
	1 		& \omega^{(d-1)}& \omega^{(d-1) \cdot 2}& \cdots & \omega^{(d-1) \cdot (d-1)} \\
															
\end{array} } \right],
\end{equation}  
where $\omega = e^{i 2 \pi /d}$. This coin is not as symmetric as the Grover coin, and we see this reflected in our analysis of quantum walk unitaries.  For a DFT coin of any dimension,  there are only two permutation matrices that leave the coin invariant.  The identity permutation gives rise to direction-preserving automorphisms, however the permutation $(1,d,d-1, \textellipsis ,3,2)$ can create joint automorphisms.  The associated coin-permutation symmetry group is hence smaller than that with the Grover coin, however $\W_1 \subsetneq \W_2$.

\item \textbf{Asymmetric coin}: Our last coin is a random unitary. To create this unitary, we generate a matrix of random complex numbers, and subsequently orthogonalize it. Here we expect $\W_1 = \W_2$.
\end{itemize}

\section{Spectral Analysis of Quantum walk unitaries}

Using the coins from the previous section, we constructed quantum walk unitaries for multiple Cayley graphs.  We consider Cayley graphs since their construction always yields $d$-regular graphs, and their automorphism groups are closely related to the groups used to construct them. In particular we look at Cayley graphs of the abelian $\mathbb{Z}_2^d$ group (hypercube of dimension $d$) and symmetric groups of different sizes.

Cayley graphs are defined using a group $G$ and a generating set $H \subseteq G$.  The elements $g_i \in G$ form the vertices of the Cayley graph, and edges exist between two vertices $g_i$ and $g_{i'}$ if $g_{i'} \cdot g_i^{-1} \in H$.  In other words, an element of $H$, say $h_j$, defines an edge between $g_i$ and $g_{i'}$ when $h_j \cdot g_i = g_{i'}$.   Thus each element of $G$ will have $\abs H$ edges, making it an $\abs H$-regular graph. $H$ is called a generating set if and only if its elements produce a Cayley graph that is connected.


The shift operator, $\hat S$, for Cayley graphs will thus be
\begin{equation}
\hat S = \sum_{i=0}^{\abs G - 1} \sum_{j=0}^{\abs H - 1} \ketbra {h_j \cdot g_i, j} {g_i, j}
\end{equation}
where $h_j \cdot g_i$ is the group composition operation between two elements $h_j$ and $g_i$.

\subsection{Hypercubes - Cayley graphs of the abelian group $\mathbb{Z}_2^d$}

\begin{figure}
\centering
\includegraphics[scale=1.25]{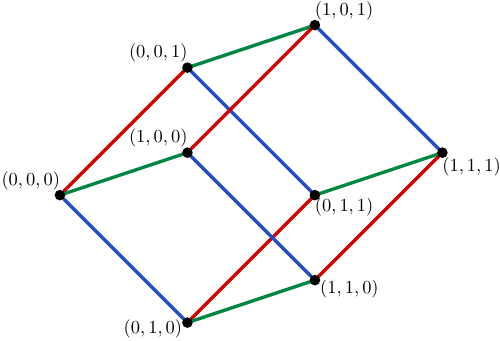}
\caption{The $3$D cube with edges colored {\color{red} red}, {\color{blue} blue} and {\color{Clover} green} to signify logical XOR with {\color{red} $001$}, {\color{blue} $010$} and {\color{Clover} $100$} respectively.
}
\label{f:hypercube}
\end{figure}

In a hypercube, vertices can be labeled using bit strings from $\mathbb{Z}_2^d$.  Hence a $d$-dimensional hypercube has a vertex Hilbert space, $\H_v$, of size $2^d$.   Two vertices labeled by bit strings $x$ and $y$ are connected if $x$ and $y$ differ in exactly one bit, that is, $$h(x \oplus y) = 1$$ where $h(a)$ is the Hamming weight of string $a$ and $x \oplus y$ is the bitwise XOR of the bit strings corresponding to vertices $x$ and $y$.  In the Cayley graph formalism, this implies the generating set $H$ contains all weight-one bit strings.  For any vertex $x$, there can only be $d$ other bit strings which satisfy the above equation, hence each vertex of the hypercube has degree $d$.  Thus, each vertex has $d$ basis states and the dimension of the coin Hilbert space $\H_c$ is also $d$.  Thus the unitary describing the evolution of states in a hypercube will be of dimension $|\H_c| \cdot |\H_v| = d \cdot 2^d$.  The $3$D cube is displayed as an example in \figref{f:hypercube}.

The hypercube is an interesting case, because it does not satisfy the sufficient condition derived in~\cite{Krovi2006a} for a graph to exhibit quantum walks with infinite hitting times.  Its automorphism group is Abelian, and all its irreducible representations have dimension 1.  However, it was observed in~\cite{Krovi2006a} that walks on the hypercube can indeed exhibit infinite hitting times with some coins.  This was presumed to be an effect of the coin symmetry. In this paper, we can now validate that intuition.

In the hypercube, the shift operator $\hat S$ can be expressed as
\begin{equation}
\hat S = \sum_{x=0}^{2^d-1} \sum_{j=0}^{d-1} \ketbra {x \oplus e_j, j} {x, j}
\end{equation}
where $e_j$ is the $n$-bit string corresponding to the number $2^j$, and the notation $\ket {a,b} \equiv \ket a \otimes \ket b$. Thus the shift operator transfers the state from vertex $x$ along edge $j$ to the vertex $x \oplus e_j$.

We analyzed the evolution unitary for hypercubes of dimension $3$, $4$ and $5$.  We choose the vertex labeled $2^d - 1$ as the final vertex with the absorbing wall.  Thus the projector onto the final vertex is $\hat \Pi_f = \ketbra {2^d - 1} {2^d - 1} \otimes \hat \Id_d$.  To avoid infinite hitting time at this final vertex, one must ensure that the initial state has no components in the IHT subspace of the unitary.  Note that the IHT subspace depends only on the choice of final vertex.

\begin{table}
\begin{tabular}{c|c|c|c}
\multicolumn{4}{c}{\textbf{$\mathbf{3}$D CUBE}} \\ 
\hline
\hline
\multicolumn{4}{c}{\textbf{Grover coin}} \\ 
\hline
\multicolumn{3}{c|}{Eigenspaces} &  \multirow{5}{*}{\parbox{2.1cm}{\centering Yes,\\ infinite hitting \\ time exists}}\\
\cline{1-3}
 $m_k$ & $\mathbf{k}$ & $|V_k|$ & \\ 
\cline{1-3}
$2$ & $\mathbf{6}$ & $3$ &  \\
$4$ & $\mathbf{3}$ & $0$ &  \\
\cline{2-3}
& $|\H| = 24$ & $|V|= 6$  & \\
\hline
\hline
\multicolumn{4}{c}{\textbf{DFT coin}} \\ 
\hline
\multicolumn{3}{c|}{Eigenspaces} &  \multirow{6}{*}{\parbox{2.1cm}{\centering Yes,\\ infinite hitting \\ time exists}}\\
\cline{1-3}
 $m_k$ & $\mathbf{k}$ & $|V_k|$ & \\ 
\cline{1-3}
$2$ & $\mathbf{2}$ & $1$ &  \\
$8$ & $\mathbf{2}$ & $0$ &  \\
$4$ & $\mathbf{1}$ & $0$ &  \\
\cline{2-3}
& $|\H| = 24$ & $|V|= 2$  & \\
\hline
\hline
\multicolumn{4}{c}{\textbf{Asymmetric coin}} \\ 
\hline
\multicolumn{3}{c|}{Eigenspaces} &  \multirow{4}{*}{\parbox{2.1cm}{\centering No, infinite hitting time does not exist }}\\
\cline{1-3}
 $m_k$ & $\mathbf{k}$ & $|V_k|$ & \\ 
\cline{1-3}
$24$ & $\mathbf{1}$ & $0$ &  \\
\cline{2-3}
& $|\H| = 24$ & $|V|= 0$  & \\
\hline
\hline
\end{tabular}
\caption{Decomposition of the evolution unitary $\hat U$ (in $\H$) for walks on the $3$D cube.  There exist $m_k$ eigenspaces of dimension $k$ each with corresponding IHT subspace $V_k$ of dimension $|V_k|$. The total IHT subspace dimension is $|V| = \sum m_k \cdot |V_k|$. }
\label{t:3Dcube}
\end{table}

\begin{table}
\begin{tabular}{c|c|c|c}
\multicolumn{4}{c}{\textbf{$\mathbf{4}$D HYPERCUBE}} \\ 
\hline
\hline
\multicolumn{4}{c}{\textbf{Grover coin}} \\ 
\hline
\multicolumn{3}{c|}{Eigenspaces} &  \multirow{6}{*}{\parbox{2.1cm}{\centering Yes,\\ infinite hitting \\ time exists}}\\
\cline{1-3}
 $m_k$ & $\mathbf{k}$ & $|V_k|$ & \\ 
\cline{1-3}
$2$ & $\mathbf{18}$ & $14$ &  \\
$2$ & $\mathbf{6}$ & $2$ &  \\
$4$ & $\mathbf{4}$ & $0$ &  \\
\cline{2-3}
& $|\H| = 64$ & $|V|= 32$  & \\
\hline
\hline
\multicolumn{4}{c}{\textbf{DFT coin}} \\ 
\hline
\multicolumn{3}{c|}{Eigenspaces} &  \multirow{6}{*}{\parbox{2.1cm}{\centering Yes,\\ infinite hitting \\ time exists}}\\
\cline{1-3}
 $m_k$ & $\mathbf{k}$ & $|V_k|$ & \\ 
\cline{1-3}
$4$ & $\mathbf{8}$ & $5$ &  \\
$4$ & $\mathbf{4}$ & $1$ &  \\
$8$ & $\mathbf{2}$ & $0$ &  \\
\cline{2-3}
& $|\H| = 64$ & $|V|= 24$  & \\
\hline
\hline
\multicolumn{4}{c}{\textbf{Asymmetric coin}} \\ 
\hline
\multicolumn{3}{c|}{Eigenspaces} &  \multirow{4}{*}{\parbox{2.1cm}{\centering No, infinite hitting time does not exist }}\\
\cline{1-3}
 $m_k$ & $\mathbf{k}$ & $|V_k|$ & \\ 
\cline{1-3}
$64$ & $\mathbf{1}$ & $0$ &  \\
\cline{2-3}
& $|\H| = 64$ & $|V|= 0$  & \\
\hline
\hline
\end{tabular}
\caption{Decomposition of the evolution unitary $\hat U$ (in $\H$) for walks on the $4$D hypercube.  There exist $m_k$ eigenspaces of dimension $k$ each with corresponding IHT subspace $V_k$ of dimension $|V_k|$. The total IHT subspace dimension is $|V| = \sum m_k \cdot |V_k|$.}
\label{t:4Dcube}
\end{table}

\begin{table}
\begin{tabular}{c|c|c|c}
\multicolumn{4}{c}{\textbf{$\mathbf{5}$D HYPERCUBE}} \\ 
\hline
\hline
\multicolumn{4}{c}{\textbf{Grover coin}} \\ 
\hline
\multicolumn{3}{c|}{Eigenspaces} &  \multirow{6}{*}{\parbox{2.1cm}{\centering Yes,\\ infinite hitting \\ time exists}}\\
\cline{1-3}
 $m_k$ & $\mathbf{k}$ & $|V_k|$ & \\ 
\cline{1-3}
$2$ & $\mathbf{50}$ & $45$ &  \\
$4$ & $\mathbf{10}$ & $5$ &  \\
$4$ & $\mathbf{5}$ & $0$ &  \\
\cline{2-3}
& $|\H| = 160$ & $|V|= 110$  & \\
\hline
\hline
\multicolumn{4}{c}{\textbf{DFT coin}} \\ 
\hline
\multicolumn{3}{c|}{Eigenspaces} &  \multirow{8}{*}{\parbox{2.1cm}{\centering Yes,\\ infinite hitting \\ time exists}}\\
\cline{1-3}
 $m_k$ & $\mathbf{k}$ & $|V_k|$ & \\ 
\cline{1-3}
$2$ & $\mathbf{10}$ & $7$ &  \\
$2$ & $\mathbf{4}$ & $2$ &  \\
$4$ & $\mathbf{2}$ & $1$ &  \\
$60$ & $\mathbf{2}$ & $0$ &  \\
$4$ & $\mathbf{1}$ & $0$ &  \\
\cline{2-3}
& $|\H| = 160$ & $|V|= 22$  & \\
\hline
\hline
\multicolumn{4}{c}{\textbf{Asymmetric coin}} \\ 
\hline
\multicolumn{3}{c|}{Eigenspaces} &  \multirow{4}{*}{\parbox{2.1cm}{\centering No, infinite hitting time does not exist }}\\
\cline{1-3}
 $m_k$ & $\mathbf{k}$ & $|V_k|$ & \\ 
\cline{1-3}
$160$ & $\mathbf{1}$ & $0$ &  \\
\cline{2-3}
& $|\H| = 160$ & $|V|= 0$  & \\
\hline
\hline
\end{tabular}
\caption{Decomposition of the evolution unitary $\hat U$ (in $\H$) for walks on the $5$D hypercube.  There exist $m_k$ eigenspaces of dimension $k$ each with corresponding IHT subspace $V_k$ of dimension $|V_k|$. The total IHT subspace dimension is $|V| = \sum m_k \cdot |V_k|$.}
\label{t:5Dcube}
\end{table}

The results of spectral decomposition of the unitaries are displayed in \tabref{t:3Dcube}, \tabref{t:4Dcube} and \tabref{t:5Dcube}.  For each of the proposed quantum walks in this paper, we ask whether the walk may potentially exhibit infinite hitting time or not.  First, we note that infinite hitting times are possible in all the tested hypercubes for quantum walks that use the Grover or DFT coin. The lack of graph symmetry is outlined by the fact that there are no degenerate eigenspaces in any of the unitaries using the asymmetric coin. As illustrated in the tables, the dimension of the IHT subspace is generally larger for quantum walks that use the Grover coin than those that use the DFT coin. 
For quantum walks that use the DFT coin, eigenspaces of dimension $<d$ may also contribute to the IHT subspace. Specifically in the $3$D cube, none of the eigenspaces are of dimension greater than three, or even equal to three, yet infinite hitting times are possible.  This points to the fact that there may be a more general condition that can predict infinite hitting times in these types of walks.

\subsection{Cayley graphs of the symmetric group}

We consider Cayley graphs of symmetric groups with different generating sets.  For the groups $S_3$ and $S_4$ we consider elements $(1,2,3)$ and $(1,2,3,4)$ respectively as the final vertices.

\begin{figure}
\centering
\includegraphics[scale=1.25]{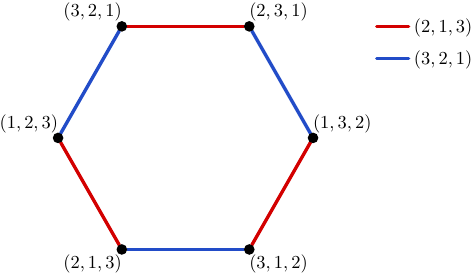}
\caption{The Cayley graph of $S_3$ with generating set $H = \{(2,1,3), (3,2,1)\}$, where application of the permutation {\color{red} $(2,1,3)$} corresponds to a {\color{red} red} edge and {\color{blue} $(3,2,1)$} corresponds to a {\color{blue} blue} edge.
}
\label{f:S32}
\end{figure}

\begin{itemize}
\item $S_3$ with generating set $H = \{(2,1,3), (3,2,1)\}$: This Cayley graph is a circle of the six elements of $S_3$, as shown in \figref{f:S32}.  Quantum walks on the circle do not exhibit infinite hitting time except when the coin operator is the identity.

\begin{figure}
\centering
\includegraphics[scale=1.25]{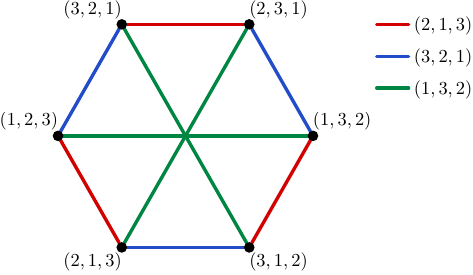}
\caption{The Cayley graph of $S_3$ with generating set $H = \{(2,1,3), (3,2,1), (1,3,2)\}$, where application of the permutation {\color{red} $(2,1,3)$} corresponds to a {\color{red} red} edge, {\color{blue} $(3,2,1)$} corresponds to a {\color{blue} blue} edge and {\color{Clover} $(1,3,2)$} corresponds to a {\color{Clover} green} edge.
}
\label{f:S33}
\end{figure}

\begin{table}
\begin{tabular}{c|c|c|c}
\multicolumn{4}{c}{\textbf{CAYLEY GRAPH OF $\mathbf{S_3}$ WITH}}\\
\multicolumn{4}{c}{$\mathbf{H = \{(1,2), (1,3), (2,3)\}}$ \footnote{$(1,3)$ implies elements $1$ and $3$ are permuted and is a shorthand for the permutation $(3,2,1)$}} \\
\hline
\hline
\multicolumn{4}{c}{\textbf{Grover coin}} \\ 
\hline
\multicolumn{3}{c|}{Eigenspaces} &  \multirow{5}{*}{\parbox{2.1cm}{\centering Yes,\\ infinite hitting \\ time exists}}\\
\cline{1-3}
 $m_k$ & $\mathbf{k}$ & $|V_k|$ & \\ 
\cline{1-3}
$2$ & $\mathbf{5}$ & $2$ &  \\
$2$ & $\mathbf{4}$ & $1$ &  \\
\cline{2-3}
& $|\H| = 18$ & $|V|= 6$  & \\
\hline
\hline
\multicolumn{4}{c}{\textbf{DFT coin}} \\ 
\hline
\multicolumn{3}{c|}{Eigenspaces} &  \multirow{6}{*}{\parbox{2.1cm}{\centering Yes,\\ infinite hitting \\ time exists}}\\
\cline{1-3}
 $m_k$ & $\mathbf{k}$ & $|V_k|$ & \\ 
\cline{1-3}
$2$ & $\mathbf{4}$ & $1$ &  \\
$4$ & $\mathbf{2}$ & $0$ &  \\
$2$ & $\mathbf{1}$ & $0$ &  \\
\cline{2-3}
& $|\H| = 18$ & $|V|= 2$  & \\
\hline
\hline
\multicolumn{4}{c}{\textbf{Asymmetric coin}} \\ 
\hline
\multicolumn{3}{c|}{Eigenspaces} &  \multirow{5}{*}{\parbox{2.1cm}{\centering No, infinite hitting time does not exist }}\\
\cline{1-3}
 $m_k$ & $\mathbf{k}$ & $|V_k|$ & \\ 
\cline{1-3}
$6$ & $\mathbf{2}$ & $0$ &  \\
$6$ & $\mathbf{1}$ & $0$ &  \\
\cline{2-3}
& $|\H| = 18$ & $|V|= 0$  & \\
\hline
\hline
\end{tabular}
\caption{Decomposition of the evolution unitary $\hat U$ (in $\H$) for walks on the Cayley graph of $S_3$ with generating set $H = \{(2,1,3), (3,2,1), (1,3,2)\}$.  There exist $m_k$ eigenspaces of dimension $k$ each with corresponding IHT subspace $V_k$ of dimension $|V_k|$. The total IHT subspace dimension is $|V| = \sum m_k \cdot |V_k|$.}
\label{t:S3_3}
\end{table}

\begin{figure*}
\centering
\includegraphics[scale=1.25]{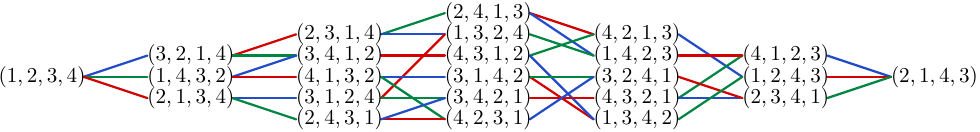}
\caption{The Cayley graph of $S_4$ with generating set $H_1 = \{(2,1,3,4), (3,2,1,4), (1,4,3,2)\}$, where application of the permutation {\color{red} $(2,1,3,4)$} corresponds to a {\color{red} red} edge, {\color{blue} $(3,2,1,4)$} corresponds to a {\color{blue} blue} edge and {\color{Clover} $(1,4,3,2)$} corresponds to a {\color{Clover} green} edge.}
\label{f:S431}
\end{figure*}

\item $S_3$ with generating set $H = \{(2,1,3), (3,2,1),$ $(1,3,2)\}$: When we add the permutation element $(1,3,2)$ to the generating set, the graph becomes more connected.  As shown in \figref{f:S33}, every vertex is now within two edges of every other vertex.  However, quantum walks on this graph can still exhibit infinite hitting times with the Grover and DFT coins!  Decomposition of the evolution unitary yields eigenspaces as in \tabref{t:S3_3}.  For the Grover coin, each of the individual subspaces have greater than three dimensions.  So they each include subspaces that have no overlap with the final vertex.  The size of the IHT subspace with the Grover coin is six. For the DFT coin, we observe a smaller IHT subspace, of dimension two.  Although quantum walks with the asymmetric coin do not actually exhibit infinite hitting times, there are degenerate eigenspaces in the evolution unitary.  This degeneracy arises from the symmetry of the graph, as explored in \cite{Krovi2006a} and \citep{Krovi2006}.

\begin{figure}
\centering
\includegraphics[scale=0.4]{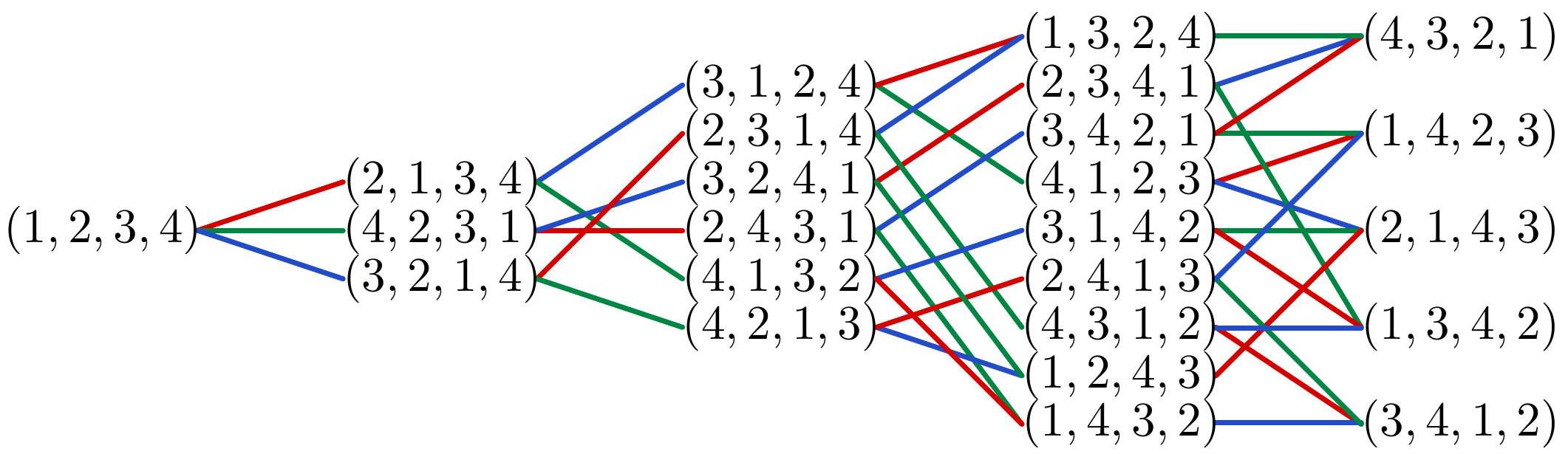}
\caption{The Cayley graph of $S_4$ with generating set $H_2 = \{(2,1,3,4), (3,2,1,4), (4,2,3,1)\}$, where application of the permutation {\color{red} $(2,1,3,4)$} corresponds to a {\color{red} red} edge, {\color{blue} $(3,2,1,4)$} corresponds to a {\color{blue} blue} edge and {\color{Clover} $(4,2,3,1)$} corresponds to a {\color{Clover} green} edge.}
\label{f:S432}
\end{figure}

\item $S_4$ with generating sets  $H_1 = \{(2,1,3,4), (3,2,1,4),$ \\ $ (1,4,3,2)\}$; $H_2 = \{(2,1,3,4), (3,2,1,4), (4,2,3,1)\}$: We discuss the Cayley graphs of $S_4$ with two generating sets of the same size to point out the differences in their spectral decomposition.  Although $H_1$ and $H_2$ contain similar elements, they produce different Cayley graphs.  $H_1$ generates a Cayley graph which can have a separation of up to six edges between two elements, as shown in \figref{f:S431}, whereas $H_2$ generates a Cayley graph that requires no more than four edges to reach every other vertex.  The Cayley graph corresponding to this generating set is displayed in \figref{f:S432}. Both graphs exhibit infinite hitting times, as shown in \tabref{t:S4_3a} and \tabref{t:S4_3b}. However, the dimension of the IHT subspace changes in each case, as shown in \tabref{t:S4_infspace}. Using the asymmetric coin, the unitary has eigenspaces of degeneracy three, which arise due to symmetries of the shift matrix, $\hat S$.

\begin{table}
\begin{tabular}{c|c|c|c}
\multicolumn{4}{c}{\textbf{CAYLEY GRAPH OF $\mathbf{S_4}$ WITH}}\\
\multicolumn{4}{c}{$\mathbf{H_1 = \{(1,2), (1,3), (2,4)\}}$} \\
\hline
\hline
\multicolumn{4}{c}{\textbf{Grover coin}} \\ 
\hline
\multicolumn{3}{c|}{Eigenspaces} &  \multirow{7}{*}{\parbox{2.1cm}{\centering Yes,\\ infinite hitting \\ time exists}}\\
\cline{1-3}
 $m_k$ & $\mathbf{k}$ & $|V_k|$ & \\ 
\cline{1-3}
$2$ & $\mathbf{14}$ & $11$ &  \\
$4$ & $\mathbf{3}$ & $1$ &  \\
$8$ & $\mathbf{3}$ & $0$ &  \\
$4$ & $\mathbf{2}$ & $0$ &  \\
\cline{2-3}
& $|\H| = 72$ & $|V|= 26$  & \\
\hline
\hline
\multicolumn{4}{c}{\textbf{DFT coin}} \\ 
\hline
\multicolumn{3}{c|}{Eigenspaces} &  \multirow{7}{*}{\parbox{2.1cm}{\centering Yes,\\ infinite hitting \\ time exists}}\\
\cline{1-3}
 $m_k$ & $\mathbf{k}$ & $|V_k|$ & \\ 
\cline{1-3}
$2$ & $\mathbf{6}$ & $5$ &  \\
$5$ & $\mathbf{3}$ & $1$ &  \\
$11$ & $\mathbf{3}$ & $0$ &  \\
$6$ & $\mathbf{2}$ & $0$ &  \\
\cline{2-3}
& $|\H| = 72$ & $|V|= 15$  & \\
\hline
\hline
\multicolumn{4}{c}{\textbf{Asymmetric coin}} \\ 
\hline
\multicolumn{3}{c|}{Eigenspaces} &  \multirow{6}{*}{\parbox{2.1cm}{\centering No, infinite hitting time does not exist }}\\
\cline{1-3}
 $m_k$ & $\mathbf{k}$ & $|V_k|$ & \\ 
\cline{1-3}
$18$ & $\mathbf{3}$ & $0$ &  \\
$6$ & $\mathbf{2}$ & $0$ &  \\
$6$ & $\mathbf{1}$ & $0$ &  \\
\cline{2-3}
& $|\H| = 72$ & $|V|= 0$  & \\
\hline
\hline
\end{tabular}
\caption{Decomposition of the evolution unitary $\hat U$ (in $\H$) for walks on the Cayley graph of $S_4$ with generating set $H_1 = {\{(2,1,3,4), (3,2,1,4), (1,4,3,2)\}}$.  There exist $m_k$ eigenspaces of dimension $k$ each with corresponding IHT subspace $V_k$ of dimension $|V_k|$. The total IHT subspace dimension is $|V| = \sum m_k \cdot |V_k|$.}
\label{t:S4_3a}
\end{table}

\begin{table}
\begin{tabular}{c|c|c|c}
\multicolumn{4}{c}{\textbf{CAYLEY GRAPH OF $\mathbf{S_4}$ WITH}}\\
\multicolumn{4}{c}{$\mathbf{H_2 = \{(1,2), (1,3), (1,4)\}}$} \\
\hline
\hline
\multicolumn{4}{c}{\textbf{Grover coin}} \\ 
\hline
\multicolumn{3}{c|}{Eigenspaces} &  \multirow{7}{*}{\parbox{2.1cm}{\centering Yes,\\ infinite hitting \\ time exists}}\\
\cline{1-3}
 $m_k$ & $\mathbf{k}$ & $|V_k|$ & \\ 
\cline{1-3}
$2$ & $\mathbf{14}$ & $11$ &  \\
$4$ & $\mathbf{6}$ & $3$ &  \\
$2$ & $\mathbf{4}$ & $1$ &  \\
$4$ & $\mathbf{3}$ & $0$ &  \\
\cline{2-3}
& $|\H| = 72$ & $|V|= 36$  & \\
\hline
\hline
\multicolumn{4}{c}{\textbf{DFT coin}} \\ 
\hline
\multicolumn{3}{c|}{Eigenspaces} &  \multirow{8}{*}{\parbox{2.1cm}{\centering Yes,\\ infinite hitting \\ time exists}}\\
\cline{1-3}
 $m_k$ & $\mathbf{k}$ & $|V_k|$ & \\ 
\cline{1-3}
$2$ & $\mathbf{4}$ & $3$ &  \\
$2$ & $\mathbf{4}$ & $1$ &  \\
$6$ & $\mathbf{3}$ & $1$ &  \\
$10$ & $\mathbf{3}$ & $0$ &  \\
$4$ & $\mathbf{2}$ & $0$ &  \\
\cline{2-3}
& $|\H| = 72$ & $|V|= 14$  & \\
\hline
\hline
\multicolumn{4}{c}{\textbf{Asymmetric coin}} \\ 
\hline
\multicolumn{3}{c|}{Eigenspaces} &  \multirow{6}{*}{\parbox{2.1cm}{\centering No, infinite hitting time does not exist }}\\
\cline{1-3}
 $m_k$ & $\mathbf{k}$ & $|V_k|$ & \\ 
\cline{1-3}
$18$ & $\mathbf{3}$ & $0$ &  \\
$6$ & $\mathbf{2}$ & $0$ &  \\
$6$ & $\mathbf{1}$ & $0$ &  \\
\cline{2-3}
& $|\H| = 72$ & $|V|= 0$  & \\
\hline
\hline
\end{tabular}
\caption{Decomposition of the evolution unitary $\hat U$ (in $\H$) for walks on the Cayley graph of $S_4$ with generating set $H_1 = {\{(2,1,3,4), (3,2,1,4), (4,2,3,1)\}}$.  There exist $m_k$ eigenspaces of dimension $k$ each with corresponding IHT subspace $V_k$ of dimension $|V_k|$. The total IHT subspace dimension is $|V| = \sum m_k \cdot |V_k|$.}
\label{t:S4_3b}
\end{table}

\begin{table}
\begin{tabular}{c|c|c}
\textbf{Generating set} & \textbf{Grover} & \textbf{DFT} \\
\hline
\hline
$H_1 = \{(1,2), (1,3), (2,4)\}$ & $26$ & $15$ \\
\hline
$H_2 = \{(1,2), (1,3), (1,4)\}$ & $36$ & $14$ \\
\hline
\end{tabular}
\caption{Differences in the dimensions of the Infinite Hitting Time (IHT) subspace in the Cayley graphs of $S_4$ with different generating sets.}
\label{t:S4_infspace}
\end{table}


\item $S_4$ with generating set $H = \{(2,1,3,4), (3,2,1,4),$ \\ $ (1,2,4,3), (1,3,2,4)\}$: Finally we also consider a generating set of size $\abs H = 4$. This set generates a graph where the maximum separation between two vertices is four edges, as displayed in \figref{f:S44}. \tabref{t:S4_4} presents the dimensions of the eigenspaces of the evolution unitary for the different coins with their corresponding IHT subspace dimensions. The dimensions of the IHT subspaces with the Grover and DFT coins are $56$ and $35$ respectively.

\begin{figure}
\centering
\includegraphics[scale=1.25]{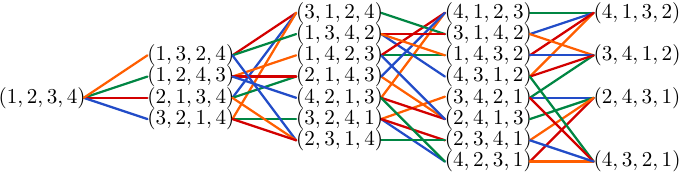}
\caption{The Cayley graph of $S_3$ with generating set $H = \{(2,1,3,4), (3,2,1,4), (1,2,4,3), (1,3,2,4)\}$, where application of the group element {\color{red} $(2,1,3,4)$} corre4ponds to a {\color{red} red} edge, {\color{blue} $(3,2,1,4)$} corresponds to a {\color{blue} blue} edge, {\color{Clover} $(1,2,4,3)$} corresponds to a {\color{Clover} green} edge and {\color{Tangerine} $(1,3,2,4)$} corresponds to a {\color{Tangerine} orange} edge.
}
\label{f:S44}
\end{figure}

\begin{table}
\begin{tabular}{c|c|c|c}
\multicolumn{4}{c}{\textbf{CAYLEY GRAPH OF $\mathbf{S_4}$ WITH}}\\
\multicolumn{4}{c}{$\mathbf{H = \{(1,2), (1,3), (3,4), (2,3)\}}$} \\
\hline
\hline
\multicolumn{4}{c}{\textbf{Grover coin}} \\ 
\hline
\multicolumn{3}{c|}{Eigenspaces} &  \multirow{7}{*}{\parbox{2.1cm}{\centering Yes,\\ infinite hitting \\ time exists}}\\
\cline{1-3}
 $m_k$ & $\mathbf{k}$ & $|V_k|$ & \\ 
\cline{1-3}
$2$ & $\mathbf{26}$ & $22$ &  \\
$2$ & $\mathbf{6}$ & $2$ &  \\
$4$ & $\mathbf{5}$ & $2$ &  \\
$4$ & $\mathbf{3}$ & $0$ &  \\
\cline{2-3}
& $|\H| = 96$ & $|V|= 56$  & \\
\hline
\hline
\multicolumn{4}{c}{\textbf{DFT coin}} \\ 
\hline
\multicolumn{3}{c|}{Eigenspaces} &  \multirow{9}{*}{\parbox{2.1cm}{\centering Yes,\\ infinite hitting \\ time exists}}\\
\cline{1-3}
 $m_k$ & $\mathbf{k}$ & $|V_k|$ & \\ 
\cline{1-3}
$2$ & $\mathbf{11}$ & $8$ &  \\
$2$ & $\mathbf{6}$ & $4$ &  \\
$2$ & $\mathbf{6}$ & $2$ &  \\
$7$ & $\mathbf{3}$ & $1$ &  \\
$7$ & $\mathbf{2}$ & $0$ &  \\
$4$ & $\mathbf{2}$ & $0$ &  \\
\cline{2-3}
& $|\H| = 96$ & $|V|= 35$  & \\
\hline
\hline
\multicolumn{4}{c}{\textbf{Asymmetric coin}} \\ 
\hline
\multicolumn{3}{c|}{Eigenspaces} &  \multirow{6}{*}{\parbox{2.1cm}{\centering No, infinite hitting time does not exist }}\\
\cline{1-3}
 $m_k$ & $\mathbf{k}$ & $|V_k|$ & \\ 
\cline{1-3}
$24$ & $\mathbf{3}$ & $0$ &  \\
$8$ & $\mathbf{2}$ & $0$ &  \\
$8$ & $\mathbf{1}$ & $0$ &  \\
\cline{2-3}
& $|\H| = 96$ & $|V|= 0$  & \\
\hline
\hline
\end{tabular}
\caption{Decomposition of the evolution unitary $\hat U$ (in $\H$) for walks on the Cayley graph of $S_4$ with generating set $H = {\{(2,1,3,4), (3,2,1,4), (1,2,4,3), (1,3,2,4)\}}$.  There exist $m_k$ eigenspaces of dimension $k$ each with corresponding IHT subspace $V_k$ of dimension $|V_k|$. The total IHT subspace dimension is $|V| = \sum m_k \cdot |V_k|$.}
\label{t:S4_4}
\end{table}

\end{itemize}

\subsection{Multiple-vertex infinite hitting time}

\begin{figure}
\centering
\includegraphics[scale = 0.17]{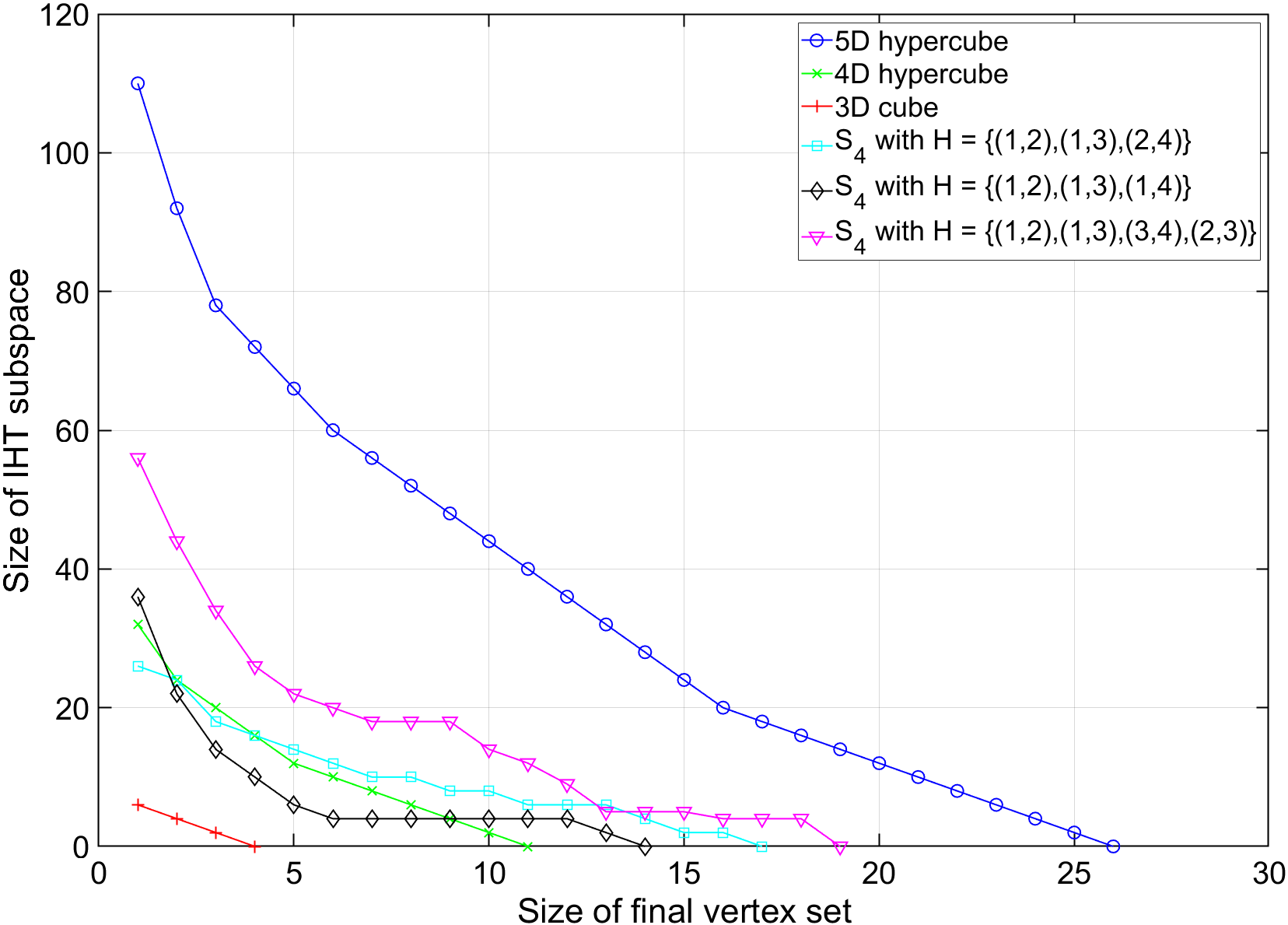}
\caption{Reduction in the dimension of the IHT subspace as  the number of vertices in the final vertex set is increased.  The evolution unitaries of six graphs with the Grover coin were analyzed.}
\label{f:graphIHTsizeVsVfCount}
\end{figure}

Until now hitting times have only been calcluated for one final vertex.  However, similar analyses can be made by considering a subset of the vertices as final vertices.  Increasing the number of final vertices leads to IHT subspaces of smaller dimension, and correspondingly the overlap between the initial state and the IHT subspace decreases. This is illustrated in \figref{f:graphIHTsizeVsVfCount}, which displays the dimension of the IHT subspace for walks using the Grover coin on graphs with different numbers of final vertices.  It is interesting to note that infinite hitting times are possible even in hypercubes of dimension three, and that infinite hitting times are possible in the 5D hypercube if even up to $25$ of the $32$ vertices are in the final vertex set.

\begin{table*}
\begin{tabular}{c||c|c|c|c|c|c|c|c}
 & $3$D & $4$D & $5$D & $S_{3,2}$ & $S_{3,3}$ & $S_{4,3}$ & $S_{4,3}$ & $S_{4,4}$\\  
 & cube & hypercube & hypercube &  &  &  &  & \\ 
\hline
$|V|$ for Grover coin & $6$ & $32$ & $110$ & $0$ & $6$ & $26$ & $36$ & $56$ \\
\hline
$|V|$ for DFT coin & $2$ & $24$ & $22$ & $0$ & $2$ & $15$ & $14$ & $35$  \\
\hline
$|V|$ for asymmetric coin & $0$ & $0$ & $0$ & $0$ & $0$ & $0$ & $0$ & $0$ \\ 
\hline
\hline
$|\mathcal{H}|$ & $24$ & $64$ & $160$ & $12$ & $18$ & $72$ & $72$ & $96$ \\
\end{tabular}
\caption{Summary of results.  The dimension of the IHT subspace $|V|$ for one final vertex is shown relative to the dimension of the unitary's Hilbert space $|\mathcal{H}|$.  Quantum walks with initial states having components in $V$ can exhibit infinite hitting time on the chosen final vertex.  $S_{i,j}$ is shorthand for a Cayley graph of the symmetric group $S_i$ with a generating set of size $j$ }
\label{t:summary}
\end{table*}

\section{Conclusion}

The objective of this paper was to explore the influence of coin symmetry on infinite hitting time. We did not find an exact relationship between the group of coin-permutation symmetries and the dimension of the IHT subspace.  However, we showed through analysis of three coins---the Grover coin, invariant under any permutation matrix; the DFT coin, invariant under only two permutation matrices; and random unitary coins lacking any symmetry---that coins with high symmetry and hence larger coin-permutation symmetry groups are associated with larger IHT subspaces.  These results are summarized in \tabref{t:summary}.  Larger IHT subspaces generally imply a larger overlap with initial states, leading to a higher probability of never reaching the final vertex.  

We also explored the effect of considering a group of vertices as ``final" vertices; a final vertex set.  It is interesting to note that these quantum walks can sill exhibit infinite hitting time, as in the case of the $5$D hypercube, which exhibits infinite hitting time with as many as $25$ vertices in the final vertex set.  Analyses of this type could be useful to determine beforehand the propensity of initial states to hit a set of target vertices in a graph. Conversely, this can also be used to choose graphs and coins to block initial states from reaching a set of target vertices.

A number of open questions remain about infinite hitting times in quantum walks. Is it possible to determine a sufficient condition for infinite hitting times using the irreducible representations of the coin-permutation symmetry group?  Can we use analyses of infinite hitting times to solve or simplify computational problems?  Are there larger automorphism groups or different types of quantum walk symmetries that more fully characterize infinite hitting times in quantum walks?

\section{Acknowledgements}

PP and TAB would like to acknowledge helpful conversations with Namit Anand, Christopher Cantwell, Yi-Hsiang Chen, Christopher Sutherland, and Paolo Zanardi. This work was supported in part by NSF Grants QIS-1719778 and FET-1911089.

\bibliographystyle{hieeetr}
\bibliography{source}

\end{document}